\documentclass[prl,twocolumn,showpacs,showkeys]{revtex4}

\usepackage{amssymb, amsmath, amsfonts, latexsym, verbatim, mathrsfs}

\usepackage{graphicx}

\begin{document}

\title{Onthological models predictively inequivalent to quantum theory}
\thanks{This work has been supported by the ARO MURI grant W911NF-11-1-0268, by NSF Grant
No. ECCS0824085, and by the Consortium for Physics of Trieste.}

\author{GianCarlo Ghirardi}
\email{ghirardi@ictp.it}
\affiliation{Department of Physics, University of Trieste, and the Abdus Salam ICTP, Trieste, Italy}
\author{Raffaele Romano}
\email{rromano@iastate.edu}
\affiliation{Department of Mathematics, Iowa State University, Ames, IA (USA) }


\begin{abstract}

\noindent Recently, it has been argued that no extension of quantum theory can have improved predictive power,
under a strong assumption of free choice of the experimental settings, and validity of quantum mechanics.
Here, under a different free choice assumption, we describe a model which violates this statement for almost all states
of a bipartite two-level system, in a possibly experimentally testable way. From consistency with quantum mechanics and
the non-signalling principle, we derive a bound on the local averages for the family of deterministic ontological
theories our model belongs to.

\end{abstract}


\keywords{Hidden variables, non-locality, free will}

\maketitle


{\it Introduction ---}
In this work we address the following question: could a theory, which is conceived as a
completion of quantum mechanics, be experimentally distinguishable from it?
By {\it completion} we mean that the theory should be consistent with
quantum mechanics, that is, it should fully reproduce all the quantum outcomes
in a suitable regime, but it could provide a more refined description of the microscopic
reality. By {\it experimentally distinguishable} we mean that there should not be physical
principles making this deeper description fully unaccessible to any observer, irrespective of
experimental complications. For instance, these are the features of classical mechanics
when compared to statistical mechanics~\footnote{Obviously, in this case the deeper description is, in principle, completely accessible.}.
In practice, the completion involves the consideration of the so-called {\it ontic state} of the system, the (in principle)
most accurate specification of its physical state. The request of the experimental testability of the completion does not
require the precise knowledge of the ontic state, but only the accessibility of some information about it, which allows
more accurate predictions than those implied by the mere knowledge of the quantum state vector.

Among others issues, the question we have raised at the beginning has been recently answered in the negative under a specific assumption of
free measurement choice~\cite{colbeck}. In this paper we exhibit a model which provides a positive answer, opening new perspectives on the
debate of the completeness of quantum mechanics. For deterministic ontological models of this type describing a pair of two-level systems,
we also derive constraints concerning how the local averages can differ from the quantum ones.

The family of completions of quantum mechanics has been usually denoted as {\it hidden variable models},
although more recently the term {\it ontological models} is preferred. In the past,
the main motivation for their introduction was the attempt to provide a description of
the micro-world as close as possible to that of the macro-world, interpreting
all (or some of) the non-classical features of quantum mechanics (as probabilism
and non-locality) as ignorance of the ontic state~\cite{harrigan}, supplying a
richer information on the state of the system than that given by the state vector of standard quantum theory.
After the theorems of Bell~\cite{bell} and Kochen-Specker~\cite{kochen}, proving that non-locality and contextuality
are unavoidable in these theories, other issues are investigated, noticeably the meaning of the quantum state,
as describing an element of reality or rather a state of knowledge~\cite{pusey,colbeck2,lewis},
the possibility to deviate from quantum mechanics~\cite{leggett,colbeck3,branciard,ghirardi},
the role of the measurement independence assumption~\cite{brans,hall},
and the dimension of the ontic state space~\cite{hardy,montina,montina2}.
We describe in more detail the aspects which are relevant for this work.

In quantum mechanics, the state of a system is represented by a vector $\psi$ in a suitable
Hilbert space. Observable quantities are represented by Hermitian operators, whose spectra contain the only
possible measurement outcomes. For general states, these outcomes are known only probabilistically.
Completeness of the theory means that it is impossible to have a more detailed description of the state
than that given by $\psi$, even in principle. In particular,
only a probabilistic knowledge of measurement outcomes is
possible. We consider a bipartite system, the separated subsystems being identified
by $A$ and $B$. The observables pertaining to these subsystems are denoted by $A(a)$,
$B(b)$, and the corresponding Hermitian operators by $\hat{A}(a)$, $\hat{B}(b)$,
where $a$ and $b$ are vectors which identify the specific observables taken into account.

In an ontological model of quantum mechanics there is a deeper specification
of the state of the system, the ontic state $\lambda$, living in a space which, for what concerns us, is not relevant to identify precisely.
For sake of simplicity we assume that $\lambda$ are continuous variables,
but our considerations apply to completely general cases.
The ontic state represents the complete description of the state of the system~\footnote{In this work,
it will be apparent whether $\lambda$ or rather $\{\psi, \lambda\}$ is considered as the ontic state of the system.},
which,  however, is not fully accessible. To each state $\psi$ is associated a
distribution $\rho_{\psi} (\lambda)$, with $\rho_{\psi} (\lambda) \geqslant 0$ and
\begin{equation}\label{norm}
    \int \rho_{\psi} (\lambda) d \lambda = 1 \qquad {\rm for \, all} \, \psi.
\end{equation}
These distributions might have overlapping supports for different $\psi$
or not~\cite{harrigan}. 
We limit our attention to
deterministic ontological models, in which the measurement outcomes are fully specified by the
ontic state (more generally, $\lambda$ could determine only their probabilities).
Contextuality implies that these outcomes can depend on the full context of measured observables;
we will then denote them as $A_{\psi}(a,b,\ldots,\lambda)$ and $B_{\psi}(a,b,\ldots,\lambda)$, depending on the subsystem
they refer to. Consistency of the ontological models with quantum mechanics means that
all the quantum averages are reproduced, in particular
\begin{equation}\label{locav}
    \int A_{\psi}(a,b,\ldots,\lambda) \rho_{\psi} (\lambda) d \lambda = \langle A(a) \rangle_{\psi},
\end{equation}
for the local averages, where $\langle A(a) \rangle_{\psi} = \langle \psi \vert \hat{A}(a) \vert \psi \rangle$
and similarly for $B$. But also quantum correlations should be obtained,
\begin{equation}\label{corr}
     \int A_{\psi}(a,b,\ldots,\lambda) B_{\psi}(a,b,\ldots,\lambda) \rho_{\psi} (\lambda) d \lambda = \langle A(a) B(b)
     \rangle_{\psi},
\end{equation}
where $\langle A(a) B(b) \rangle_{\psi} = \langle \psi \vert \hat{A}(a) \otimes \hat{B}(b) \vert \psi \rangle$.

Notice that the ontic state $\lambda$ is necessarily not fully accessible. In fact, in a theory like the one we are devising,
it is just the ignorance of the precise value of $\lambda$ which cancels non-locality in the averages (\ref{locav}), in accordance
with the fact that quantum mechanics does not allow faster-than-light communication. Nonetheless, at least in principle, one can
investigate whether it is possible to avoid superluminal communication while taking advantage of some information on $\lambda$.
It is exactly this property which must characterize a completion of quantum theory which could be experimentally distinguishable
from it, the crucial problem addressed in this paper.

With this in mind, suppose that $\lambda$ is equivalently described by two variables $(\mu, \tau)$,
where $\mu$ denotes the unaccessible part, i.e. the one which must be averaged over, and $\tau$ the accessible one.
Knowledge of $\tau$ should not allow superluminal communication; therefore, by writing $\rho_{\psi} (\lambda) = \rho_{\psi, \tau} (\mu) \rho_{\psi} (\tau)$, we can compute
\begin{eqnarray}\label{locavcnlhv}
    \int A_{\psi}(a,b,\ldots,\lambda) \rho_{\psi, \tau}(\mu) d \mu &=& f_{\psi}(a, \tau), \nonumber \\
    \int B_{\psi}(a,b,\ldots,\lambda) \rho_{\psi, \tau}(\mu) d \mu &=& g_{\psi}(b, \tau),
\end{eqnarray}
which are the local averages of the theory at the intermediate level. They express the {\it non-signalling conditions} in our context.
It remains true that the theory reproduces quantum mechanics when $\tau$ is averaged over,
\begin{eqnarray}\label{locavcnlhv2}
    \int f_{\psi}(a, \tau) \rho_{\psi}(\tau) d \tau &=& \langle A(a) \rangle_{\psi}, \nonumber \\
    \int g_{\psi}(b, \tau) \rho_{\psi}(\tau) d \tau &=& \langle B(b) \rangle_{\psi},
\end{eqnarray}
but, if the state vector $\psi$ is supplemented with the information on $\tau$, the theory
could be experimentally distinguishable from quantum mechanics~\footnote{We assume that $\tau$
could be prepared at the source, and then possibly communicated to $A$ and $B$. But different scenarios are possible, for instance, the accessible and unaccessible parts of the ontic state could be contextual, that is, they could depend on $a$ and $b$.}.
Theories having this structure have been initially introduced by A. Leggett~\cite{leggett}, and then further analyzed in the case of
maximally entangled states~\cite{colbeck3,branciard}. Here we provide one model fitting this class, describing arbitrarily entangled states.

{\it A model with non trivial local averages ---}
The model is a generalization to arbitrary states of the famous Bell's model for the
singlet state of a pair of two-level systems,
and its details can be found elsewhere~\cite{ghirardi3}. Here we review
only the main results using a different notation, which is more convenient for the present purposes. An arbitrary state $\psi$ is written as
\begin{equation}\label{genstate}
    \vert \psi \rangle = \sin{\frac{\theta}{2}} \vert 00 \rangle + \cos{\frac{\theta}{2}} \vert 11 \rangle,
\end{equation}
with $\theta \in [0, \pi/2]$; if $\theta = 0$, $\vert \psi \rangle$
is a separable state state; if $\theta = \pi/2$ it is a maximally entangled state. The hidden variable $\lambda$ is a unit vector in
the 3-dimensional real space and the pair $\{\psi,\lambda\}$ is identified with the ontic state. We consider local observables represented by the operators $\hat{A} (a) = \sigma \cdot a$ and $\hat{B} (b) = \sigma \cdot b$, where $a$ and $b$ are real, unit vectors and $\sigma = (\sigma_x, \sigma_y,\sigma_z)$ is the vector of Pauli matrices. In particular, $\sigma_z$ is defined so that $\vert 0 \rangle$ and $\vert 1 \rangle$ are its $+1$ and $-1$ eigenvectors respectively. For sake of simplicity we assume that $a$ and $b$ lie in the plane orthogonal to the direction of propagation of the entangled particles~\footnote{Nonetheless, the model can be built in the general case of arbitrary directions $a$ and $b$, and non necessarily traceless local operators.}. The possible outcomes are in the set $\{ -1, 1 \}$, and they are defined as
\begin{equation}\label{defouta}
    A_{\psi} (a, b, \lambda) = \left\{
                                    \begin{array}{ll}
                                      +1, & \hbox{\rm if $\hat{a} \cdot \lambda \geqslant \cos{\xi}$,} \\
                                      -1, & \hbox{\rm if $\hat{a} \cdot \lambda < \cos{\xi}$,}
                                    \end{array}
                                  \right.
\end{equation}
and
\begin{equation}\label{defoutb}
    B_{\psi} (b, \lambda) = \left\{
                                    \begin{array}{ll}
                                      +1, & \hbox{\rm if $b \cdot \lambda \geqslant \cos{\chi}$,} \\
                                      -1, & \hbox{\rm if $b \cdot \lambda < \cos{\chi}$.}
                                    \end{array}
                                  \right.
\end{equation}
In the previous relations, $\hat{a} = \hat{a} (a, b)$ is in the plane of $a$ and $b$, as detailed in~\cite{ghirardi3}; moreover,
$\cos{\xi} = - \langle A(a) \rangle_{\psi}$, and $\cos{\chi} = - \langle B(b) \rangle_{\psi}$.

In~\cite{ghirardi3} it has been proved that this model is predictively equivalent to quantum mechanics
when $\lambda$ is uniformly distributed on the unit sphere, $\rho (\lambda) = 1/4 \pi$. At this point, we express $\lambda$ by using spherical coordinates $(\mu, \tau)$, whose pole is identified by the direction of the incoming particle; $\mu$ is the azimuthal angle and $\tau$ the polar angle.
They represent the unaccessible and the accessible part of $\lambda$, respectively. In (\ref{locavcnlhv}) we have $\rho_{\psi, \tau} (\mu) = 1 / 2 \pi$, and, by construction, integration over $\mu$ cancels non-locality in local averages. We find that
\begin{equation}\label{result}
  f_{\psi} (a, \tau) = \frac{1}{\pi} \, \cos^{-1}{\left(\frac{2 \langle A(a) \rangle_{\psi}^2}{\sin^2{\tau}} - 1 \right)} - 1,
\end{equation}
if $\vert \tau - \frac{\pi}{2} \vert \leqslant \xi$ and $f_{\psi} (a, \tau) = - 1$ otherwise,
and a similar relation (with $\xi$ replaced by $\chi$) for $g_{\psi} (b, \tau)$,
which clearly shows that in general $f_{\psi} (a, \tau)$ and $g_{\psi} (b, \tau)$ differ from the local quantum averages.
While this toy-model is completely artificial, it provides evidence that models compatible with quantum mechanics, but in principle distinguishable from it, are indeed possible,
with measurement settings $a$ and $b$ freely chosen.

{\it On the local averages of any ontological theory ---}
The requirements of equivalence with quantum mechanics and the non-signalling conditions put severe constraints on the probabilities at the intermediate level. In the following, we derive a bound on the local averages of any deterministic ontological model for quantum mechanics describing a pair of 2-level systems arbitrarily entangled, at the intermediate level. We adopt the aforementioned description of the ontic state in terms of the state vector, $\psi$, of an additional accessible part, $\tau$, and a non-accessible one, $\mu$, and assume that this splitting is independent of the local settings $a$ and $b$.

\begin{figure}
  \includegraphics[width=6cm]{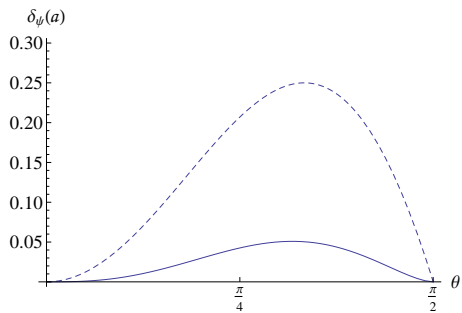}
  \caption{Constraint for $\delta_{\psi} (a)$ as a function of entanglement, parameterized by $\theta$ (dashed line), when $\hat{A} (a) = \sigma_z$. The solid line represents $\delta_{\psi} (a)$ for the generalization of the Bell's model. The bound can be saturated in the special case of the local observable here considered, by redefining the outcomes supports.}\label{fig2}
\end{figure}

While our derivation holds for both subsystems, for sake of simplicity we focus on the $A$ part and its corresponding local average $f_{\psi} (a, \tau)$. As a measure of the distance between the local averages at the intermediate level of the ontological theory and quantum mechanics we introduce the quantity
\begin{equation}\label{distance}
    \delta_{\psi} (a) = \int \Big( f_{\psi} (a, \tau) - \langle A (a) \rangle_{\psi} \Big)^2 \, \rho_{\psi}(\tau) d \tau,
\end{equation}
which is the variance of the variable $f_{\psi} (a, \tau)$ over the distribution $\rho_{\psi}(\tau)$.
We consider a generic observable $A(a)$, represented in quantum mechanics by $\hat{A} (a)$ with eigenvalues $\pm 1$. By elementary manipulations of (\ref{distance}), and
considering that $f_{\psi} (a, \tau) \in [-1, 1]$, we can write
\begin{equation}\label{ref}
  \delta_{\psi} (a) \leqslant \int \vert f_{\psi} (a, \tau) \vert \rho_{\psi} (\tau) d \tau - \langle A (a) \rangle_{\psi}^2.
\end{equation}
We now focus on the integral in the r.h.s. of this inequality. Since $\hat{A} (-a) = - \hat{A} (a)$, it follows that $A_{\psi} (-a) = - A_{\psi} (a)$, and then
$f_{\psi} (-a, \tau) = - f_{\psi} (a, \tau)$. Consider $2n + 1$ unit vectors $\gamma_j$, $j = 0, 1, \ldots, 2n$ such that $\gamma_0 = a$ and $\gamma_{2n} = -a$.
Further assume that $2n$ pairs of local measurements are performed according to the following scheme:
(i) when $j$ is even, the local observables are given by $A (\gamma_j)$ and $B (\gamma_{j + 1})$;
(ii) when $j$ is odd, the local observables are given by $A (\gamma_{j + 1})$ and $B (\gamma_{j})$.
The measurement outcomes for $A (\gamma_j)$ and $B (\gamma_j)$ are $\pm 1$. Therefore, by using that
\begin{eqnarray}\label{step0}
    &&\vert A_{\psi}(\gamma_j, \gamma_{j + 1}, \lambda) - B_{\psi}(\gamma_j, \gamma_{j + 1}, \lambda) \vert = \\
    && \qquad 1 - A_{\psi}(\gamma_j, \gamma_{j + 1}, \lambda) B_{\psi}(\gamma_j, \gamma_{j + 1}, \lambda), \nonumber
\end{eqnarray}
after multiplication of both sides by $\rho_{\psi, \tau} ({\mu})$ and integration over $\mu$, it follows that
\begin{equation}\label{step1}
    \vert f_{\psi} (\gamma_j, \tau) - g_{\psi} (\gamma_{j + 1}, \tau) \vert \leqslant
    1 - E_{\psi, \tau} (\gamma_j, \gamma_{j + 1}),
\end{equation}
where we have defined the correlation at the intermediate level as
\begin{equation}\label{corrint}
    E_{\psi, \tau} (a, b) = \int A_{\psi} (a, b, \lambda) B_{\psi} (a, b, \lambda) \rho_{\psi, \tau} (\mu) d \mu
\end{equation}
When $j$ is odd, a similar argument leads to a relation analogous to (\ref{step1}), with $\gamma_j$ and $\gamma_{j + 1}$ exchanged.
By summing all these relations, and considering that $f_{\psi} (-a, \tau) = - f_{\psi} (a, \tau)$, we find that
\begin{eqnarray}\label{step3}
    \vert f_{\psi} (a, \tau) \vert \leqslant n &-& \frac{1}{2} \sum_{k = 0}^{n - 1} \Big( E_{\psi, \tau} (\gamma_{2k}, \gamma_{2k+1}) + \nonumber \\
    && \quad + E_{\psi, \tau} (\gamma_{2k+2}, \gamma_{2k+1}) \Big).
\end{eqnarray}
By further multiplying by $\rho_{\psi} (\tau)$ and integrating over $\tau$,
\begin{equation}\label{finalrel}
    \int \vert f_{\psi} (a, \tau) \vert \rho_{\psi} (\tau) d \tau \leqslant \min_{\gamma_1, \ldots, \gamma_n; \, n \in \mathbb{N}} \Omega_{\psi} (a, n),
\end{equation}
where we have defined
\begin{eqnarray}\label{omegan}
    \Omega_{\psi} (a, n) = n &-& \frac{1}{2} \sum_{k = 0}^{n - 1} \Big( \langle A (\gamma_{2k}) B (\gamma_{2k + 1}) \rangle_{\psi} + \nonumber \\
    && \quad + \langle A (\gamma_{2k + 2}) B (\gamma_{2k + 1}) \rangle_{\psi}
    \Big),
\end{eqnarray}
and the minimum is taken by arbitrarily varying the vectors $\gamma_1, \ldots, \gamma_{2n - 1}$, for any $n$.
Therefore, by taking into account the joint correlations arising in an arbitrary number of measurements, we can write
\begin{equation}\label{constraint2}
    \delta_{\psi} (a) \leqslant \min_{\gamma_1, \ldots, \gamma_n; \, n \in \mathbb{N}} \Omega_{\psi} (a, n) - \langle A (a) \rangle_{\psi}^2.
\end{equation}
Notice that, when $\psi$ is a maximally entangled state, both terms in the r.h.s. of (\ref{constraint2}) vanish, and then $f_{\psi} (a, \tau) = \langle A(a) \rangle_{\psi}$
for all $a$ (a rigorous proof, using analogous arguments, can be found in~\cite{ghirardi}). For arbitrary states, it is in general difficult to find the minimum in (\ref{constraint2}), therefore
we have resorted to a numerical analysis, providing strong evidence that this minimum is given by $\cos{\theta}$.
Therefore, we conjecture that (\ref{constraint2}) can be simply written as
\begin{equation}\label{constraint3}
    \delta_{\psi} (a) \leqslant \cos{\theta} - \langle A (a) \rangle_{\psi}^2.
\end{equation}
In Fig. \ref{fig2} we show how this constraint for $\delta_{\psi} (a)$ varies with entanglement, and plot the
corresponding curve for the aforementioned model.

{\it Optimality of the bound ---}
Our generalization of the Bell's model does not saturate the bound on the local average at the
intermediate level and this rises the question whether this constraint is optimal or not. In
general it is not, because in (\ref{ref}) we have used the fact that $f_{\psi} (a, \tau)^2 \leqslant
\vert f_{\psi} (a, \tau) \vert$, which is a strict inequality unless $f_{\psi} (a, \tau) \in \{-1, 0, 1\}$.
However, we can provide a model which saturates the bound when the local observable is $\sigma_z$. This is done by modifying
our model, such that the supports on the unit sphere, where the two outcomes $A_{\psi}$ and $B_{\psi}$
take the values $+ 1$ and $- 1$, remain consistent with (\ref{locav}), (\ref{corr}) and (\ref{locavcnlhv}). We define
$A_{\psi} (a, b, \lambda) = - {\rm sign} \langle A (a) \rangle_{\psi}$ if $\vert \mu - \mu_a \vert \leqslant \pi$ and $\vert \tau -
\frac{\pi}{2} \vert \leqslant \eta$, and $A_{\psi} (a, b, \lambda) = {\rm sign} \langle A (a) \rangle_{\psi}$ otherwise, where
$\mu_a$ is the azimuthal angle of $a$, $\sin{\eta} = 1 - \vert \langle A(a) \rangle_{\psi} \vert$,
and $\tilde{a} = \tilde{a} (a, b)$ plays the same role of $\hat{a}$
previously defined. Its explicit expression is not relevant here. Similarly, $B_{\psi} (b, \lambda) = - {\rm sign} \langle B (b) \rangle_{\psi}$
if $\vert \mu - \mu_b \vert \leqslant \pi$ and $\vert \tau - \frac{\pi}{2} \vert \leqslant \zeta$, and $B_{\psi} (b, \lambda) = {\rm sign} \langle
B (b) \rangle_{\psi}$ otherwise, where $\mu_b$ is the azimuthal angle of $b$ and $\sin{\zeta} = 1 - \vert \langle B(b) \rangle_{\psi} \vert$. We find that $f_{\psi} (a, \tau) = 0$ for $\vert \tau - \frac{\pi}{2} \vert \leqslant \eta$, and  $f_{\psi} (a, \tau) = {\rm sign} \langle A (a) \rangle_{\psi}$ otherwise, leading to
\begin{equation}\label{optimal}
  \delta_{\psi} (a) = \vert \langle A(a) \rangle_{\psi} \vert - \langle A(a) \rangle_{\psi}^2,
\end{equation}
which reproduces (\ref{constraint3}) when $\sigma_z$ is measured.

It turns out that the bound (\ref{constraint3}) is optimal for an arbitrary local measurement
when the state $\psi$ is factorized, that is, $\theta = 0$. Also in this case, this can be proven
by redistributing the outcomes domains on the unit sphere, in particular by defining $A_{\psi}
(a, \lambda) = - {\rm sign} \langle A (a) \rangle_{\psi}$ if $\vert \tau - \frac{\pi}{2} \vert
\leqslant \eta^{\prime}$, and $A_{\psi} (a, \lambda) = {\rm sign} \langle A (a) \rangle_{\psi}$
otherwise, where $\sin{\eta} = 2 \sin{\eta^{\prime}}$, and $B_{\psi} (b, \lambda)$ is defined in order
to reproduce the quantum averages. The model is local, and the non-signalling conditions are automatically
satisfied. Therefore, (\ref{constraint3}) expresses the optimal bound for $\delta_{\psi}(a)$ in the extremal
cases of maximal or null entanglement, for generic local measurements.

{\it Conclusions ---}
In this work we have proven that ontological models of quantum theory which are compatible
with it, but possibly experimentally distinguishable from it, are possible. In our model,
the standard free will assumption, which involves the measurements of $A$ and $B$, is satisfied,
and superluminal communication is impossible. Our result seems to contradict a recent result implying
that quantum mechanics is maximally informative~\cite{colbeck}, but this is not the case, since
in our model we do not rely on the same mathematical expression of the free will assumption.
Moreover, our model
provides the first example of a crypto-nonlocal theory in which the local averages differ from the
quantum mechanical ones for arbitrary non-maximally entangled states of a pair of qubits, and it
is consistent with former results on maximally entangled states~\cite{colbeck3,branciard,ghirardi,dilorenzo}.
It proves that the local part of the hidden variables can be nontrivial, and suggests that possible
deviations from quantum mechanics on the local averages could be observed in the case of non-maximally entangled
states.

Finally, we have derived an explicit upper bound on the local averages
of any deterministic ontological theory for quantum mechanics, when
the system is given by a pair of qubits. This constraint is determined
by the requirements that: (i) the theory respects the non-signalling condition,
when one takes into account the accessible part of $\lambda$, and (ii) it is
compatible with quantum mechanics, that is, its predictions are the standard
ones when the full average over $\lambda$ is performed. In the case of general
local measurements, our bound is optimal only for factorized or maximally entangled
states. For arbitrary entanglement, it is optimal when $\hat{A} (a) = \sigma_z$.


\end{document}